In-Plane Orbital Texture Switch at the Dirac Point in the Topological Insulator Bi$_2$Se$_3$

Yue Cao[1], J. A. Waugh[1], X.-W. Zhang[2,3], J.-W. Luo[3], Q. Wang[1], T. J. Reber[1], S. K. Mo[4], Z. Xu[5], A. Yang[5], J. Schneeloch[5], G. Gu[5], M. Brahlek[6], N. Bansal[6], S. Oh[6], A. Zunger[7], Daniel S. Dessau[1]

1 Department of Physics, University of Colorado, Boulder, Colorado 80309, USA
2 Department of Physics, Colorado School of Mines, Golden, Colorado 80401, USA
3 National Renewable Energy Laboratory, Golden, Colorado 80401, USA
4 Advanced Light Source, Lawrence Berkeley National Lab, Berkeley, California 94720, USA
5 Condensed Matter Physics and Materials Science Department, Brookhaven National Laboratory, Upton, New York 11973, USA
6 Department of Physics and Astronomy, Rutgers University, Piscataway, New Jersey 08854, USA
7 University of Colorado, Boulder, Colorado 80309, USA

Topological insulators are novel macroscopic quantum-mechanical phase of matter [1, 2], which hold promise for realizing some of the most exotic particles [3, 4] in physics as well as application towards spintronics and quantum computation [1, 2, 3, 5, 6]. In all the known topological insulators, strong spin-orbit coupling [5, 6] is critical for the generation of the "protected" massless surface states [7, 8, 9, 10]. Consequently, a complete description of the Dirac state should include both the spin and orbital (spatial) parts of the wavefunction. For the family of materials with a single Dirac cone, theories [11, 12, 13] and experiments [14, 15] agree qualitatively, showing the topological state has a chiral spin texture that changes handedness across the Dirac point (DP), but they differ quantitatively on how the spin is polarized. Limited existing theoretical ideas predict chiral local orbital angular momentum on the two sides of the DP [16]. However, there have been neither direct measurements nor calculations identifying the global symmetry of the spatial wavefunction. In this paper we present the first results from angle-resolved photoemission experiment and first-principles calculation that both show, counter to current predictions, the in-plane orbital wavefunctions for the surface states of Bi$_2$Se$_3$ are *asymmetric* relative to the DP, switching from being tangential to the k-space constant energy surfaces above DP, to being radial to them below the DP. Because the orbital texture switch occurs exactly at the DP this effect should be intrinsic to the topological physics, constituting an essential yet missing aspect in the description of the topological Dirac state. Our results also indicate that the spin texture may be more complex than previously reported, helping to reconcile earlier conflicting spin resolved measurements [14, 15].

While intensely studied [11, 12, 13, 14, 15], spin is likely not a good quantum number of the Dirac state. Instead we may wish to consider whether the total angular momentum **J**, which includes the spin and orbital aspects of the wavefunctions, could be the good quantum number [16]. Thus the orbital properties of the total wavefunction may be highly relevant for describing the topological surface states of these materials, though this possibility has been essentially unexplored. Also, while the existence of the Dirac state is determined by tracking the time reversal symmetry of the bulk bands [17, 18, 19], its effective description could be constructed by considering the time reversal and real space symmetries of the surface state itself, without information from the bulk bands [8, 12, 20]. To investigate which features of the Dirac state the "effective model" captures we need to check the both the spin and orbital properties of the electronic wavefunction. These questions motivate us to explore the orbital aspects of the Dirac states.

$Bi_2Se_3$ is among an established class of topological insulators with a single Dirac cone at Γ. As shown in Ref [8, 12, 20], the bulk and Dirac bands are all made by the out-of-plane $p_z$ orbitals, since the in-plane $p_x$ and $p_y$ orbitals are well-separated from the $p_z$ orbitals by strong crystal field effects. However, in the Dirac states, the strong spin-orbit coupling explicitly mixes the in-plane states with the $p_z$ orbital. We will focus on the in-plane states here, as they show new and unexpected physics.

Within the prevailing description of the Dirac state [8, 12, 16, 20], we expect (see Supplementary Online Materials) that the in-plane part of the orbital wavefunction in the vicinity of the Dirac point is symmetric relative to the Dirac Point. As we will show, this is counter to the experimental observation.

Angle-Resolved Photoemission Spectroscopy (ARPES) provides a unique opportunity to directly measure k-state orbital structures with different symmetries, via the photon polarization selection rules [21] (called "the Matrix Element Effect"). The measured ARPES intensity $I \propto \left|\langle \psi_f | A \cdot p | \psi_i \rangle\right|^2 \propto \left|\langle \psi_f | E \cdot r | \psi_i \rangle\right|^2$ where **A** is the electromagnetic gauge and **p** is the electron momentum. $|\psi_i\rangle$ and $|\psi_f\rangle$ represent the initial state of the electron in the solid and the final state of the photo-excited electron, respectively. In the presence of a linearly polarized electric field, only electronic states $|\psi_i\rangle$ of a particular symmetry will contribute to the measured ARPES intensity. This is because the matrix element includes integration across all spatial dimensions, so if the parity of the product $\langle \psi_f | A \cdot p | \psi_i \rangle$ is overall odd with respect to a particular mirror plane, the integration goes to zero and the ARPES intensity vanishes. By properly arranging the experimental geometry, it is often possible to adjust the parity of $|\psi_f\rangle$ and $|A \cdot p|$ such that the ARPES intensity vanishes in a certain direction, thus determining the parity (symmetry) of the initial

state wavefunction $|\psi_i\rangle$. This is a very powerful method to determine the symmetry of the initial state orbitals or wavefunctions, and is the goal of the present experiment.

As discussed above, to a very good approximation we only need to consider the p-like wavefunctions [8, 12, 20] for the Dirac states and bulk bands nearest to the DP. In the experiment, the incident photons come at a glancing angle ~ 7 degrees to the sample plane and can have either p-polarization (photon electric field vector, drawn with yellow arrow, in the orange-coloured scattering plane) or s-polarization (E field perpendicular to the scattering plane). These possibilities are illustrated in FIG. 1 a. In both configurations, *only* the electron analyzer is rotated to collect data, so that the relative angles between the sample coordinate axes and the photon beam coordinates (polarization and Poynting or incident vector) remain *unchanged*. Detailed information about the ARPES set up and data taking could be found in the Supplementary Online Materials.

The **E** field of p-polarization points out-of-plane. Therefore it leads to a strong ARPES cross-section for the out-of-plane $p_z$ orbitals and a weak cross section for the in-plane orbitals. In contrast, s-polarization data has a strong cross-section for the in-plane orbitals and a weak cross-section for the out-of-plane orbitals. This is observed in FIGs. 1 b and c, which show the energy-momentum intensity plot along Γ-K taken with s and p-polarization, respectively. The ARPES intensity of the Dirac cone using p-polarization is 2~3 times stronger than using s-polarization, confirming the Dirac states have a large $p_z$ component and non-negligible contribution from in-plane states. In addition to the surface states that make up the Dirac cone, the bulk valence band can also be observed in the interior of the Dirac cone below the DP using p-polarization. This indicates that the bulk valence band has a major $p_z$ component, as shown in Ref [8, 20].

FIG. 2 a shows constant energy surfaces (CESs) through the Dirac cone for different energies relative to the DP (left to right) and for both polarizations (s and p as marked on the right of the panel). The bottom row shows data from p-polarization, mainly made up of the $p_z$ states. These are seen to be almost uniform around the constant energy surfaces for all energy cuts. In contrast, the data taken with s-polarization has drastic intensity changes around the constant energy surfaces. In particular the data above the DP (left 2 columns) both show vanishing spectral weight parallel to the electric field, while the data below the DP (right 2 columns) shows suppressed spectral weight normal to the electric field. To determine whether this weight distribution is related to a specific crystalline orientation (sample frame) or relative to the photon field (lab frame), we rotated the sample crystalline axes in multiple 5-degree steps of the angle $\phi$ (FIG. 1 a) about the sample normal while keeping all other experimental parameters the same. These data, shown in FIG. 2 b from left to right columns are almost identical with sample rotation, illustrating that this pattern is not due to any particular arrangement relative to the 6-fold crystalline axes but is more general. We found that if we take similar spectra with deeper binding energies below DP (not shown), we start to see the six-fold symmetric feature away from Γ, associated with the crystalline axes, and the pattern rotates with the sample. These portions, associated with the bulk spectra, are distinct from the surface state components that we focus on here.

We now use a symmetry analysis across various mirror planes to disentangle the symmetries of the various in-plane states that contribute, requiring us to only consider s-polarized photons. A helpful mirror plane to consider is the one defined by the sample surface normal and the photon Poynting vector (orange plane in FIG. 1a, in the $k_x$ direction in the lab frame, and shown as the orange lines in FIG. 2 c). Relative to this $k_x$-$k_z$ lab-based mirror plane the s-polarized photon field **E** has an odd parity, while it has an even parity relative to the green $k_y$-$k_z$ mirror plane. The free-electron final state $|\psi_f\rangle$ is even with respect to both these mirror planes. As labeled on FIG. 2 c, this constrains the initial state wavefunctions $|\psi_i\rangle$ to have a certain parity with respect to these mirror planes, so that the ARPES intensity will vanish in the correct symmetry locations if the overall parity of the matrix element is odd. Above the DP, the in-plane states along the green G-$k_y$ line, and thus the initial state $|\psi_i\rangle$, must have odd symmetry with respect to this mirror plane (FIG. 2 c, top) for a zero matrix element. Similarly, for the in-plane states below the DP, there is vanishing weight along the orange G-$k_x$ line and so the initial state is even with respect to this mirror plane (FIG. 2 c, bottom). Along $k_x$ above DP and along $k_y$ below the DP, there is strong spectral weight, the matrix elements are overall even, and the initial state parities can similarly be determined. Note neither of these mirror planes are necessarily along any of the high symmetry crystalline directions of the sample, as is seen from the data of FIG. 2 b.

We could deduce from FIG. 2 c the in-plane p orbitals $|\psi_i\rangle$ that are consistent with the symmetry constraints discussed above, that is, odd with respect to the orange and green mirror planes above the DP and even with respect to these planes below the DP. As can be seen, these orbital wavefunctions are tangential to the constant energy surface above the DP and then switch to being radial to the constant energy surface below the DP. This is shown more clearly in FIG. 3 as the orange orbitals, while showing the out-of-plane $p_z$ components of the wavefunction in green. It is quite evident from symmetry analysis that linear polarization cleanly disentangles ARPES intensity contributions from different p orbitals (see Supplementary Online Materials for more details), while the circular polarization used in previous experiments [22, 23, 24, 25] focused on the spin chirality or the "handedness" of the wavefunctions.

The measured orbital texture is captured in our first-principles calculations based on the local density approximation. The calculated Dirac surface bands, the bulk conduction band, and the bulk valence band nearest to the DP are drawn in FIG. 4 a. In FIG. 4 b we show the calculated component of the atomic $p_y$ orbitals of Dirac cone state on constant energy surfaces above and below the DP, each summing over a 20 meV energy window. The calculation well reproduces the experimental measurement of the orbital texture of the Dirac state. For example, above the DP, the calculated $p_y$ component maximizes along the lab axis $k_x$ and minimized along $k_y$ (where the $p_x$ orbital dominates the in-plane states). Similar to the experiment, we also found that rotating the sample axes relative to the lab frame by angle $\phi$ has minimal effect on the calculated $p_y$ intensity distribution, especially near the DP. We discuss more details of the first-principles calculation in Methods and in the Supplementary Online Materials.

To further trace the origin of the orbital texture switch of the Dirac state, we would like to zoom in close to the Dirac point as this is the region where the Dirac physics is unaffected by lattice effects or hybridization to the bulk bands. Due to the limit of experimental resolution we explore this near-DP region using the calculated orbital densities. The calculation for each constant energy surface is displayed as a function of sample azimuth angle (as defined in FIG 4 b) in FIG 4 c. Most notable is the switch of the intensity distribution above and below the DP. Also as the energy gets closer to the DP, the intensity variation fits better to a $\sin 2\theta$ or $\cos 2\theta$ distribution. At 300 meV above the DP, the intensity curve shows a six-fold modulation on top of the $\cos 2\theta$ function. This additional modulation might come from the hybridization to the bulk bands, or from the non-isotropic term of spin-orbit coupling [26]. The switch of the intensity distribution above and below the DP is a signature of the orbital texture switch. To identify whether the switch is sudden or gradual, we define the Orbital Polarization Ratio (OPR) $\lambda$

$$\lambda(\Delta E) = \frac{I_0(\Delta E) - I_{90}(\Delta E)}{I_0(\Delta E) + I_{90}(\Delta E)}$$

where $I_\theta$ is the intensity of the matrix element around the CES, with the 0 degree angle in $\theta$ defined in FIG. 4 b (note that this angle is relative to the lab frame, while the angle $\phi$ is between the sample and lab frame – see FIGs. 1 and 2). It is seen from FIG. 4 d the $\lambda$ linearly approaches zero as the energy approaches the DP, and to within the statistical error, changes sign exactly at the DP. Thus while the orbital is almost purely polarized to tangential 200meV above DP and radial 200meV below DP as seen in the experiments, at the DP it is an equal superposition of the tangential and radial states. Our observation that the orbital texture switch happens exactly at the DP provides strong evidence that the orbital polarization switch is an intrinsic feature of the Dirac surface state. This new behavior indicates the unexpected richness of the surface states of topological insulators and calls for a more complete description to take into account these discoveries.

We would like to point out the key differences of the in-plane orbital texture between this work and previous predictions. First, our work shows that the orbital wavefunction of the Dirac states is strongly energy dependent. Using the model of Ref [8, 20], the Dirac state wavefunction is essentially independent of energy and momentum. Second, the in-plane orbitals are asymmetric relative to the DP, as compared to the model. Third, the in-plane states, once away from the Dirac point, are not an equal superposition of the $p_x$ / $p_y$ orbitals. A complete effective model of the topological state should include these features.

The current work may also help reconcile conflicting spin resolved measurements [14, 15]. Since different electric field configurations (including the ratio of the in-plane vs. out-of-plane electric fields) relative to the sample pick up different orbital wavefunctions, and electron spin and orbital are strongly coupled, it is natural that different experimental geometries could give different spin polarizations. Actually the asymmetric orbital wavefunction suggests spin polarization may also be asymmetric across the DP, which should be explored by future theories and experiments.

**Methods**

The ARPES experiments were performed at Beamline 10.0.1 of the Advanced Light Source, LBL. Data has been taken both from $Bi_2Se_3$ thin films and bulk samples and have shown consistent results. The $Bi_2Se_3$ thin films were prepared using a two-step growth method, as describe in Ref [27] and the Supplementary Online Materials, and were protected with a Se overlayer after growth, and decapped *in situ* by heating in the final vacuum environment of the analysis chamber. The bulk samples were cleaved *in situ* at 50K with a base pressure better than $5 \times 10^{-11}$ Torr.

We performed calculations of a 6 quintuple-layer slab of $Bi_2Se_3$. The orbital character of electronic states is obtained by projecting the calculated plane-wave based wavefunctions $|\psi_{nk}\rangle$ onto spherical harmonics $|J_l^{R_i} Y_{lm}^{R_i}\rangle$ including p-orbitals ($l = 1$) centered at the position of the ions $R_i$.

$$|\psi_{nk}\rangle = \sum_i^{N_{at}} \sum_{lm} \alpha_{m,nk}^{R_i} |J_l^{R_i} Y_{lm}^{R_i}\rangle$$

Also we choose an s-like ($l = 0$) final state to represent the electron state photon-excited to vacuum, so that it is always even with respect to the proposed mirror planes. The projection strategy is so chosen to bring about the symmetry information in the atomic orbitals, without contributions from any special mirror plane. We find that the Dirac cone states especially those away from the DP are a hybridization of topological surface states and bulk states as evidenced by their wavefunction distribution. This feature of the topological states was found in 2D HgTe / CdTe topological insulator by some of us [28]. In order to distinguish the pure topological surface states from bulk band states, we projected the states layer-by-layer and summed up the contribution of each atomic orbital with an exponential weighting away from the surface, and we confirmed the choice of decay distance does not affect the calculation qualitatively. The results shown in FIG. 4 b have a decay distance of 0.5nm, which approximates the ARPES probe depth.

**Acknowledgements**
We acknowledge helpful discussions with S.-C. Zhang, S.-R. Park, M. Hermele, A. Essin and G. Chen. The ARPES work was performed at the Advanced Light Source, LBL, which is supported by the DOE, Office of Basic Energy Sciences.

**Figure 1**. **ARPES energy-momentum intensity plots at the Γ point for s and p photon polarizations.** **a** shows the experimental configuration, where the sample frame is shown in red and the lab frame (which contains the electron detector) in blue. The sample axes can be rotated by the angle θ relative to the lab frame, though the normal of the sample and lab frame always stay aligned. The incident photon beam makes an angle of ~7 degrees relative to the lab (and sample) planes and has varying polarizations ranging from full s (**E** field parallel to the sample plane) to full p (**E** field in the orange $k_x$-$k_z$ plane). **b** and **c** are ARPES cuts along the Γ-K direction of the sample taken with s and p polarization, respectively, with the sample Γ-K axis lying in the $k_x$ lab frame direction.

**Figure 2**. **Deducing the orbital texture from the constant energy surface intensity plots. a** The experimental constant energy surface intensity plot at different energies relative to the DP. The Γ-K direction of the sample is parallel to the lab $k_x$. **b** s-polarization experimental constant energy surface intensity plots as a function of in-plane sample orientation and energies relative to the DP, in 5 degree steps, from left to right columns. The axes shown are the crystal axes that are attached to the lattice orientation. The lab frame direction is unchanged as the sample is rotated, and all energy labels are relative to the Dirac point. **c** Symmetry analysis of p orbitals across the orange $k_x$-$k_z$ and green $k_y$-$k_z$ lab-

frame mirror planes (see also FIG. 1 **a**). The free electron final state $|\psi_f\rangle$ is even in all cases, while the s-polarized photon field $|A \cdot p|$ is odd with respect to the orange $k_x$-$k_z$ mirror plane and even with respect to the green $k_y$-$k_z$ mirror plane. As illustrated, this gives strong or weak intensity for initial states $|\psi_i\rangle$ with symmetry that have the total matrix element even or odd, respectively. Since this result is independent of sample rotation (panel **b**), this result implies an in-plane tangential orbital texture above the DP and a radial one below (see FIG. 3).

**Figure 3. Sketch of the orbital texture switch deduced from the experimental and theoretical matrix elements. a** The green orbitals are of $p_z$ character and are viewed with p-polarized light; **b** The orange orbitals represent the in-plane $p_x$ / $p_y$ components of the orbital wavefunction and are summed so that the net in-plane wavefunctions are tangential to the constant energy surface above the DP and radial below the DP.

**Figure 4. Orbital Polarization Ratio λ switches sign at the DP. a** the Dirac surface bands (red), the bulk conduction band (green) and bulk valence band (blue) nearest to the DP, as calculated from first principles, only showing the first quadrant. **b** Calculated matrix element at different energies relative to the DP, each summed over a window of 20 meV relative to the central energy shown on the plot. **c** For each energy relative to the DP, we plot the calculated projected matrix element as a function of the sample in-plane azimuth angle (for the definition of the azimuth angle, see panel **b**, with 0 degree as marked). The dashed lines are the selected cos2θ / sin2θ fits to the calculated matrix elements shown in solid lines. **d** calculated orbital polarization ratio λ as a function of the energy relative to the DP. Note l is positive when E>$E_D$ and negative when E<$E_D$. l switches sign exactly at the DP.

Figure 1

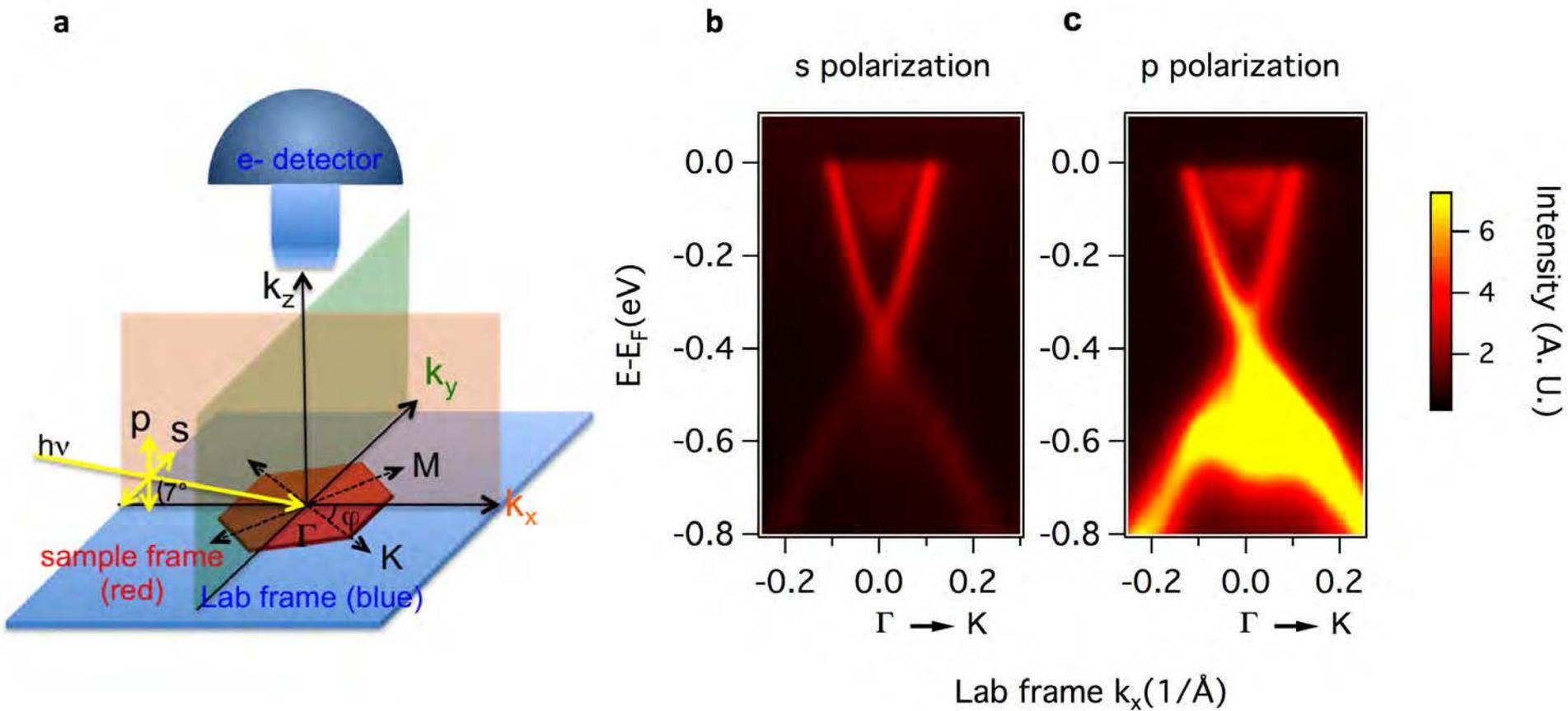

Figure 2

Figure 3

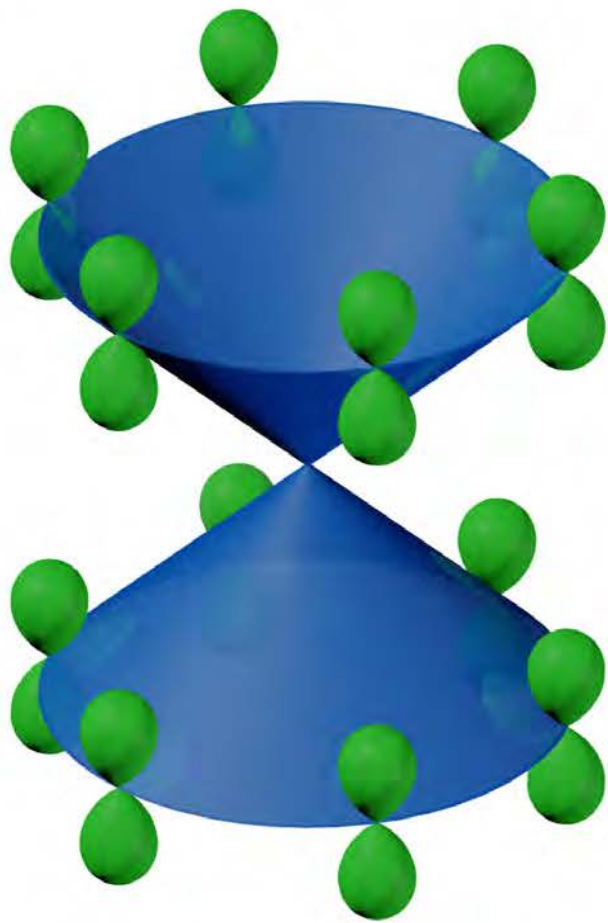 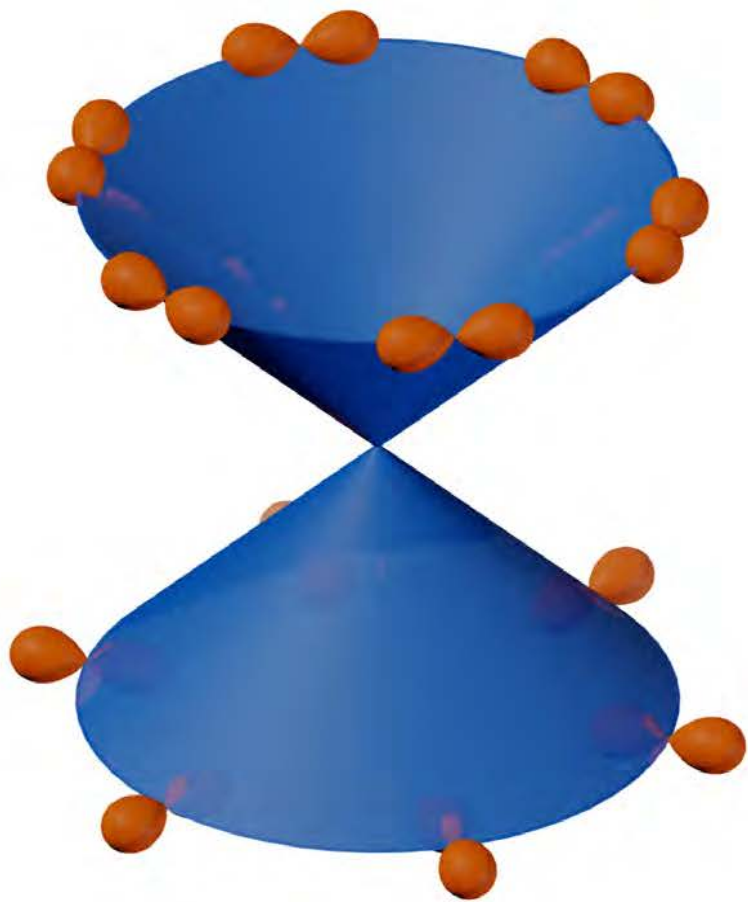

Figure 4

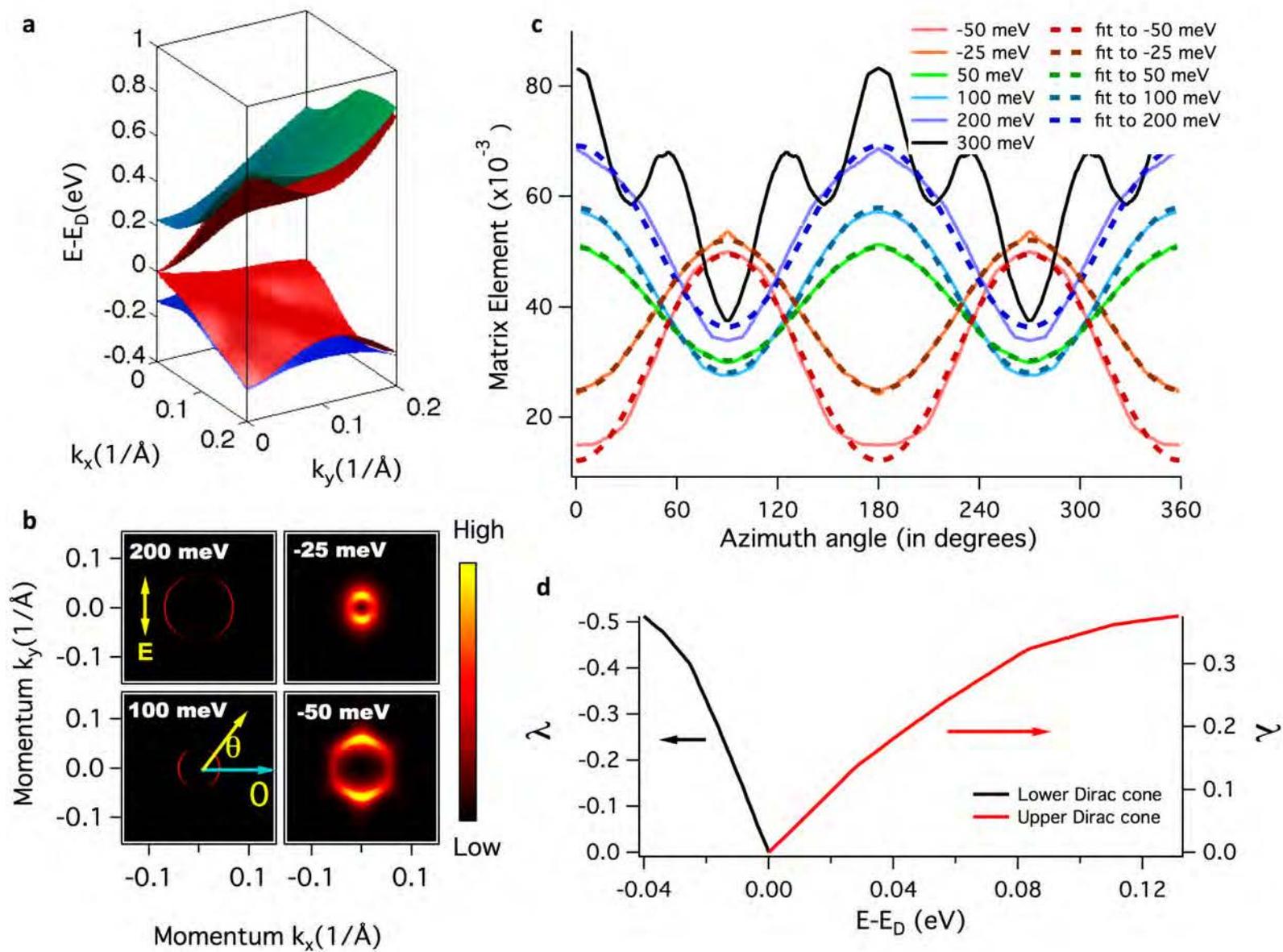

In-Plane Orbital Texture Switch at the Dirac Point in the Topological Insulator Bi$_2$Se$_3$

Supplementary Online Materials

Yue Cao[1], J. A. Waugh[1], X.-W. Zhang[2,3], J.-W. Luo[3], Q. Wang[1], T. J. Reber[1], S. K. Mo[4], Z. Xu[5], A. Yang[5], J. Schneeloch[5], G. Gu[5], M. Brahlek[6], N. Bansal[6], S. Oh[6], A. Zunger[7], Daniel S. Dessau[1]

1 Department of Physics, University of Colorado, Boulder, Colorado 80309, USA
2 Department of Physics, Colorado School of Mines, Golden, Colorado 80401, USA
3 National Renewable Energy Laboratory, Golden, Colorado 80401, USA
4 Advanced Light Source, Lawrence Berkeley National Lab, Berkeley, California 94720, USA
5 Condensed Matter Physics and Materials Science Department, Brookhaven National Laboratory, Upton, New York 11973, USA
6 Department of Physics and Astronomy, Rutgers University, Piscataway, New Jersey 08854, USA
7 University of Colorado, Boulder, Colorado 80309, USA

1. **Properties of the Dirac State Derived from Model Hamiltonians**

In this section, we would like to propose the general properties of the Dirac state derived from a general linear spin-orbit coupling Hamiltonian. To describe the linear dispersion of the Dirac state across the Dirac point, we expect the leading term in the Hamiltonian to be the linear coupling between electron spin and momentum, as used in Ref [1, 2, 3]. Thus

$$H = \sum_{i=x,y}\sum_{j=x,y} k_i \sigma_j$$

Of course we could put further constraint on the Hamiltonian, by considering the rotational and inversion symmetries.

**Proposition**: If the wavefunction $|k,s\rangle$ is an eigenfunction of the model Hamiltonian, with an eigenvalue of E>0, then $|k,s'\rangle \equiv \sigma_z |k,s\rangle$ is also an eigenfunction of the model Hamiltonian, with an eigenvalue of –E.

This is because

$$\sigma_z H(k,\sigma)\sigma_z = -H(k,\sigma)$$

Since the Dirac state with momentum k and energy –E<0 is singly degenerate, this indicates $|k,s'\rangle \equiv \sigma_z |k,s\rangle$ is the *only* eigenfunction with the eigenvalue -E. In the known literature [1, 2, 3]

the eigenfunction could be expanded in the complete basis set of $\left|k_{i=x,y,z}\right\rangle \otimes \left|s^z\right\rangle$. Thus the Pauli matrix will only have an effect on the spin part of the wavefunction. This further suggests for a pair of Dirac states with the same momentum k but opposite energy eigenvalues, the spatial wavefunction will be identical.

2. **Sample Information**

The orbital texture has been measured both from the $Bi_2Se_3$ bulk sample and thin films, with consistent results obtained from both the films and crystals. The bulk $Bi_2Se_3$ crystals were grown at Brookhaven National Laboratory by the unidirectional solidification method and a melting zone growth method. The films were grown at Rutgers University. Here are more details about the film growth:

High-quality $Bi_2Se_3$ films [4] were grown on $Al_2O_3$ (0001) substrate in a custom-designed SVTA MOS-V-2 MBE system (FIG. S1); the base pressure of the system was lower than $5 \times 10^{-10}$ Torr. Bi and Se fluxes were provided from Knudsen cells; the fluxes were measured using a quartz crystal microbalance, Inficon BDS-250, XTC/3. To remove traces of organic compounds from the surface, the $Al_2O_3$ (0001) substrates were exposed to UV light for 5 minutes before mounting into the growth chamber. The substrates were then heated in vacuum to a temperature of 700° C in an oxygen atmosphere of $1 \times 10^{-6}$ Torr for ~10 minutes. $Bi_2Se_3$ films of various thicknesses were then grown using the two-temperature growth process. Evolution of the film surface during growth was monitored by RHEED. After deposition of 3 QL of $Bi_2Se_3$ at 110°C, a sharp streaky pattern was observed, indicating single-crystal $Bi_2Se_3$ structure. The film was then slowly annealed to a temperature of 220°C, which helped further crystallization of the film as seen by the brightening of the specular spot. The diffraction pattern and the Kikuchi lines became increasingly sharp on further $Bi_2Se_3$ deposition. This shows that the grown films have atomically flat morphology and high crystallinity. The film quality was further improved by annealing the sample at 220°C for an hour after the growth. This process led to high quality single crystalline films with the largest terraces, highest bulk mobilities, and lowest volume carrier densities. The atomically flat terraces observed by atomic force microscopy (AFM) in FIG. S2 are much larger than any previous reports on $Bi_2Se_3$ thin films, representing the high quality of these samples.

After the growth of the films was completed, the samples were cooled down to room temperature. At ~35-45°C, Selenium gets deposited as an amorphous layer (as observed from RHEED), and we deposit ~50-60nm of Se as a capping layer. This capping layer protects the surfaces during transport to the ARPES setup, where the cap can be subsequently removed by an ~ 200°C anneal in vacuum.

3. **ARPES set up**

All the samples are measured at BL 10.0.1 of the Advanced Light Source, LBL. Data were collected with a Scienta R4000 analyzer using multiple photon energies from 37eV to 67eV. The total resolution is kept at 20-35meV. The data was taken in the Angle 14 mode with a 0.5 mm analyzer entrance slit.

The bulk samples were cleaved in situ at 50K with a base pressure better than $5\times10^{-11}$ Torr. The films were heated and decapped in the preparation chamber, with a base pressure better than $1\times10^{-9}$ during decapping and then transferred into the analysis chamber while maintaining ultra-high vacuum conditions throughout. The end station has a beam spot size of 150mm by 100mm. Since the beam comes in at a glancing angle, the actual beam spot size on the sample is 150mm by 800μm~1mm. It has been hard in general to get rid of flakes for the bulk sample; so all data presented in this paper came from the decapped $Bi_2Se_3$ film.

4. **A Brief Review of Matrix Element Effects**

The measured ARPES intensity $I \propto |\langle \psi_f | A \cdot p | \psi_i \rangle|^2 \propto |\langle \psi_f | E \cdot r | \psi_i \rangle|^2$ where **A** is the electromagnetic gauge and **p** is the electron momentum. $|\psi_i\rangle$ and $|\psi_f\rangle$ are the wavefunctions of the initial state of the electron in the solid and the final state of the photo-excited electron, respectively. As shown in FIG. 1 (A), if we choose the mirror plane to contain both the plane of the outcoming photoelectron and the sample normal, the final state $|\psi_f\rangle$ is *generally* even with respect to the mirror plane. For a non-vanishing matrix element, the $|A \cdot p|$ and $|\psi_i\rangle$ should both be even or both be odd relative to the mirror plane so that the overall parity of the matrix element is even. In Table S1, we list all non-vanishing matrix element for the possible combination of polarization and the final state of the photoelectrons.

Here we additionally show that the data we are showing does not reply on any special property of the final state wavefunction $|\psi_f\rangle$. This is confirmed by varying the incident photon energy so that different final states are chosen. The constant energy surface intensity plot 200 meV above the Dirac point for different photon energies are shown in FIG. S3.

**5. Calculating the band structure and Matrix Element via First-principles Calculation**

A. Calculation method

We used the Perdew-Burke-Ernzerhof (PBE) exchange-correlation functional [5, 6] as implemented in the Vienna ab-initio simulation package (VASP) [7], the projector-augmented wave (PAW) pseudopotential [8], and energy-cutoff of 316 eV, for computing the electronic structure of the bulk and thin film. Spin-orbit coupling (SOC) was invoked in our calculations. Since previous structural studies of bulk $Bi_2Se_3$ reported agreement with experimental lattice constants [9], we choose to relax only the internal parameters with an error of $10^{-4}$ eV/f.u. and fix the lattice constants to the experimental values for the bulk, and use a non-orthogonal k-point mesh along the reciprocal lattice vectors of 9×9×9 in the first Brillouin zone. The $Bi_2Se_3$ thin film studied here consists of 6 quintuple layers (30 atomic layers) separated by vacuum with the same thickness of the film. We considered the relaxation of atomic positions in the z direction since a vacuum along z-direction is present. The total energy of thin film was assumed to have converged when the z-component of the Hellman-Feynman force is smaller than 0.025 eV/Å. We used a k-mesh size of 6×6×1 on the surface. The calculated band structure is as shown in FIG. S4 as well as FIG. 4 (A) of the main text.

B. Surface state decomposition.

The dipole moment matrix element measured can be approximated by $|\langle f|E \cdot r|\psi_{n\bar{k}}\rangle|^2$. The initial state $|\psi_{n\bar{k}}\rangle$ of each band at each k-point can be approximately decomposed into spd Bessel functions and spherical harmonics that are non zero within spheres of a radius $R_i^W$ around each ion $R_i$. The spd and site projected wave function character of $|\psi_{n\bar{k}}\rangle$ is calculated as:

$$a_{lm,n\bar{k}}^{R_i} = \langle J_l^{R_i} Y_{lm}^{R_i} | \psi_{n\bar{k}} \rangle$$

$$= \int_{in\_sphere} d\bar{r} J_l^{R_i}(r_i) Y_{lm}^{R_i}(\theta_i, \varphi_i) \psi_{n\bar{k}}(\bar{r}) = \int_0^{R_i^W} dr_i r_i^2 \int_0^{\pi} \sin\theta_i \, d\theta_i \int_0^{2\pi} d\varphi_i J_l^{R_i}(r_i) Y_{lm}^{R_i}(\theta_i, \varphi_i) \psi_{n\bar{k}}(r_i, \theta_i, \varphi_i)$$

Thus the initial state $|\psi_{n\bar{k}}\rangle$ can be written as:

$$|\psi_{n\bar{k}}\rangle = \sum_i^{N_{at}} \sum_{lm} a_{lm,n\bar{k}}^{R_i} |J_l^{R_i} Y_{lm}^{R_i}\rangle + residual\_states,$$

The final state $|f\rangle$ can also be decomposed as:

$$|f\rangle = \sum_i^{N_{at}} \sum_{lm} b_{lm}^{R_i} |J_l^{R_i} Y_{lm}^{R_i}\rangle + residual\_states\_f.$$

We assume that the final state $|f\rangle$ is nearly homogeneously distributed in space.

Thus $|f\rangle$ is not angular dependent inside each sphere, i.e. s-like containing only s spherical harmonics around each ion $|J_0^{R_i} Y_{00}^{R_i}\rangle$ (having the symmetry as the atomic $|s\rangle$ state of ion i, and we denote $|J_0^{R_i} Y_{00}^{R_i}\rangle$ also as $|s^{R_i}\rangle$ later), that is:

$$|f\rangle = \sum_i^{N_{at}} b_{00}^{R_i} |J_0^{R_i} Y_{00}^{R_i}\rangle + residual\_states\_f.$$

Since the $|J_l^{R_i} Y_{lm}^{R_i}\rangle$ functions are normalized inside each sphere around each ion, we have $b_{00}^{R_i} = b_{00}$.

Then the matrix element $\langle f | E \cdot r | \psi_{n\bar{k}} \rangle$ can be calculated as:

$$\langle f | E \cdot r | \psi_{n\bar{k}} \rangle = b_{00} \sum_j^{N_{at}} \langle J_0^{R_j} Y_{00}^{R_j} | E \cdot r \sum_i^{N_{at}} \sum_{lm} a_{lm,n\bar{k}}^{R_i} | J_l^{R_i} Y_{lm}^{R_i} \rangle,$$

considering that $|J_l^{R_i} Y_{lm}^{R_i}\rangle$ is zero outside the sphere ($R_i^W$) around ion i, we have

$$\langle f | E \cdot r | \psi_{n\bar{k}} \rangle = b_{00} \sum_i^{N_{at}} \sum_{lm} a_{lm,n\bar{k}}^{R_i} \langle J_0^{R_i} Y_{00}^{R_i} | E \cdot r | J_l^{R_i} Y_{lm}^{R_i} \rangle.$$

We can represent the $|J_1^{R_i} Y_{11}^{R_i}\rangle$, $|J_1^{R_i} Y_{10}^{R_i}\rangle$ and $|J_1^{R_i} Y_{1-1}^{R_i}\rangle$ in $|p_x^{R_i}\rangle$, $|p_y^{R_i}\rangle$ and $|p_z^{R_i}\rangle$ that have the symmetry as the atomic $|p_x\rangle$, $|p_y\rangle$ and $|p_z\rangle$ states of ion i:

$$|J_1^{R_i} Y_{11}^{R_i}\rangle = -\frac{|p_x^{R_i}\rangle + i|p_y^{R_i}\rangle}{\sqrt{2}},$$

$$\left|J_1^{R_i}Y_{1-1}^{R_i}\right\rangle = \frac{\left|p_x^{R_i}\right\rangle - i\left|p_y^{R_i}\right\rangle}{\sqrt{2}},$$

$$\left|J_1^{R_i}Y_{10}^{R_i}\right\rangle = \left|p_z^{R_i}\right\rangle.$$

Similarly, $\left|J_2^{R_i}Y_{2m}^{R_i}\right\rangle$ ($\left|J_3^{R_i}Y_{3m}^{R_i}\right\rangle$) can be represented in d-like states $\left|d^{R_i}\right\rangle$ (f-like states $\left|f^{R_i}\right\rangle$).

For each ion, $\left\langle s^{R_i}\left|E\cdot r\right|d^{R_i}\right\rangle = 0$ if $E\cdot r$ is odd, and $\left\langle s^{R_i}\left|E\cdot r\right|f^{R_i}\right\rangle \cong 0$ since f-states are very far away from band edge that we are considering, and we denote $\left\langle s^{R_i}\left|E_x x\right|p_x^{R_i}\right\rangle = \left\langle s^{R_i}\left|E_y y\right|p_y^{R_i}\right\rangle = P$ for the x-polarized and y-polarized light.

Then for x-polarized light:

$$\left\langle f\left|E_x x\right|\psi_{n\vec{k}}\right\rangle = b_{00}P\sum_i^{N_{at}}\frac{1}{\sqrt{2}}(a_{1-1,n\vec{k}}^{R_i} - a_{11,n\vec{k}}^{R_i}),$$

and for y-polarized light:

$$\left\langle f\left|E_y y\right|\psi_{n\vec{k}}\right\rangle = b_{00}P\sum_i^{N_{at}}\frac{i}{\sqrt{2}}(a_{1-1,n\vec{k}}^{R_i} + a_{11,n\vec{k}}^{R_i}).$$

The calculated matrix element for the given energies relative to the Dirac point is shown in the main text. We have also performed calculations where the sample is rotated by 15 degrees and the intensity distribution remains unchanged. This is displayed in FIG. S5, which is the analog of the experimental data of FIG. 2 (B) of the main paper.

**Figure S1. RHEED images showing the steps of Bi$_2$Se$_3$ growth on sapphire substrates.** a, Sapphire substrate mounted in the UHV growth chamber after UV-cleaned for 5 min. b, On heating to 700 $^o$C in an O$_2$ pressure of 1x10$^{-6}$ Torr for 10 min. c, After deposition of 3 QL of Bi$_2$Se$_3$ film at 110 $^o$C. d, Specular beam spot gets brighter on annealing the film to 220 $^o$C. e, RHEED pattern gets much brighter and sharper on subsequent growth of another 29 QL at 220 $^o$C. f, Final RHEED pattern of the 32 QL film after being annealed at 220 $^o$C for an hour.

**Figure S2. 1.5 × 1.5 μm$^2$ scanned AFM image of a 300 QL thick Bi$_2$Se$_3$ film grown on Al$_2$O$_3$ (0001).** Large terraces (largest ever reported for Bi$_2$Se$_3$ thin films) are observed, further verifying the high quality of the grown films.

**Figure S3. The constant surface intensity variation of the Dirac state as a function of photon energy.** The plot shows the intensity 200 meV above the Dirac point. The observed intensity variation does not change with photon energy, indicating the final states in the matrix element are free electron states.

**Figure S4. The calculated band structure from a 6 quintuple layer slab of $Bi_2Se_3$.** All energies are relative to the Dirac point. The Dirac surface band is represented in the solid green line. The red and blue solid lines are the bulk conduction bands and bulk valence bands, respectively.

**Fig S5. Calculated constant energy surfaces of a variety of energies for two different sample orientations.** Top panels show original sample orientation with ΓK along $k_x$, and bottom panels have a sample rotation of 15° rotated counterclockwise from the top panels, while leaving the photon polarization fixed to the lab frame. The lack of any change with rotation is consistent with the experimental observations of FIG. 2 b of the main text.

**Table S1. Initial state orbital wavefunctions with non-vanishing ARPES spectral weight.** The table is shown across two key mirror planes (see Fig 1 of the main text for relevant mirror planes). For * marked states, the spectral weight is much smaller with the momentum / photon polarization combination than the one without the *; but they are not symmetry forbidden.

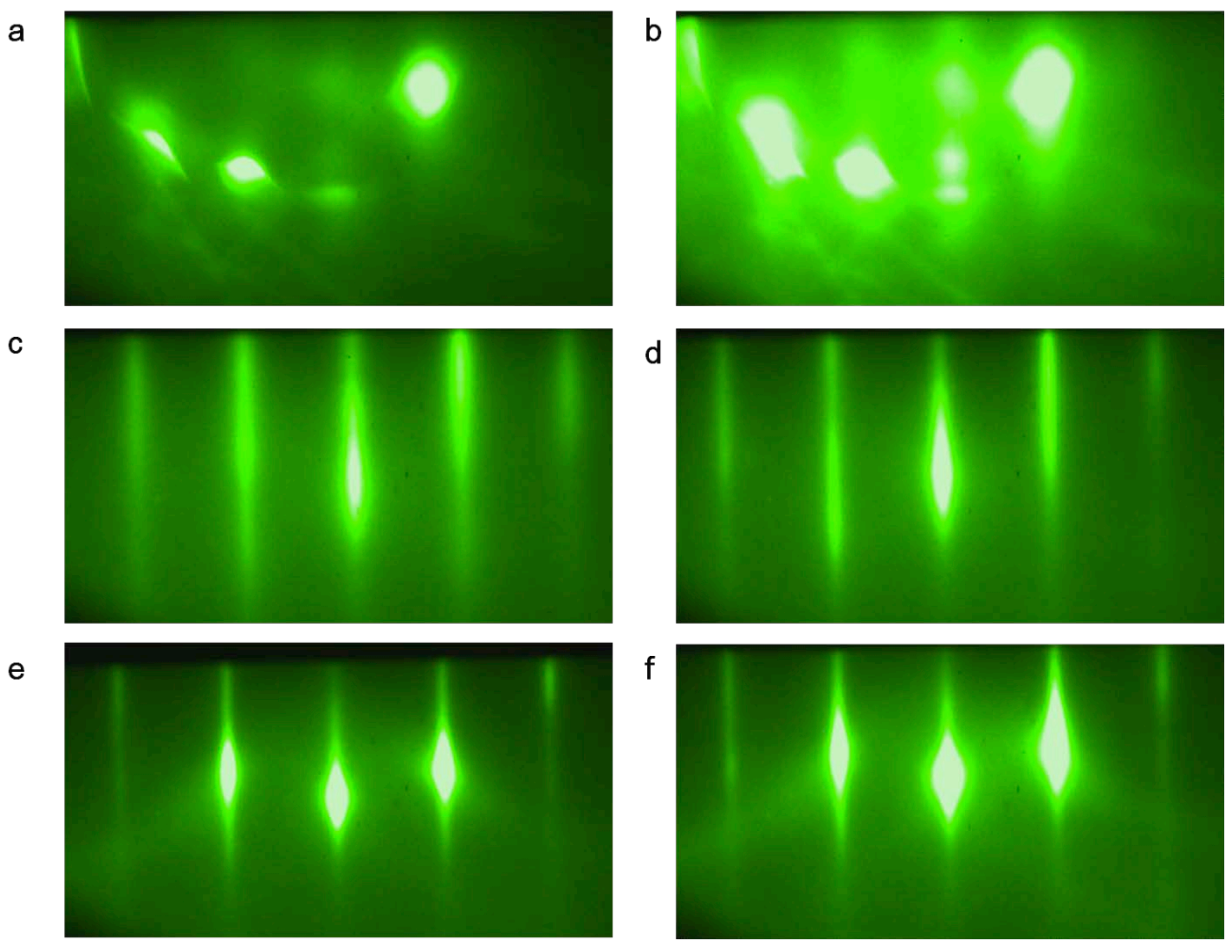

**Figure S1. RHEED images showing the steps of Bi$_2$Se$_3$ growth on sapphire substrates.** a, Sapphire substrate mounted in the UHV growth chamber after UV-cleaned for 5 min. b, On heating to 700 °C in an O$_2$ pressure of 1x10$^{-6}$ Torr for 10 min. c, After deposition of 3 QL of Bi$_2$Se$_3$ film at 110 °C. d, Specular beam spot gets brighter on annealing the film to 220 °C. e, RHEED pattern gets much brighter and sharper on subsequent growth of another 29 QL at 220 °C. f, Final RHEED pattern of the 32 QL film after being annealed at 220 °C for an hour.

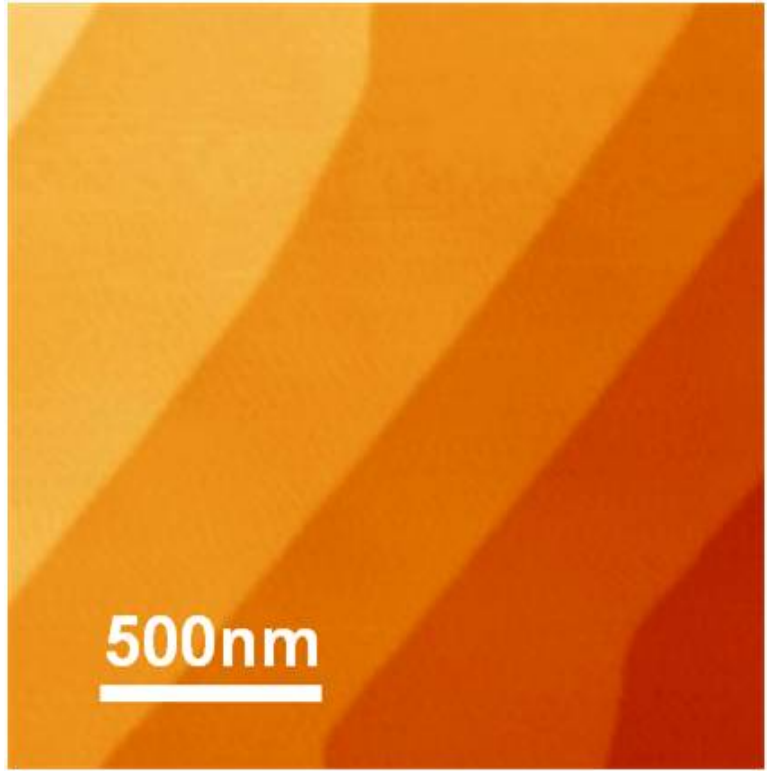

**Figure S2. 1.5 × 1.5 μm² scanned AFM image of a 300 QL thick $Bi_2Se_3$ film grown on $Al_2O_3$ (0001).** Large terraces (largest ever reported for $Bi_2Se_3$ thin films) are observed, further verifying the high quality of the grown films.

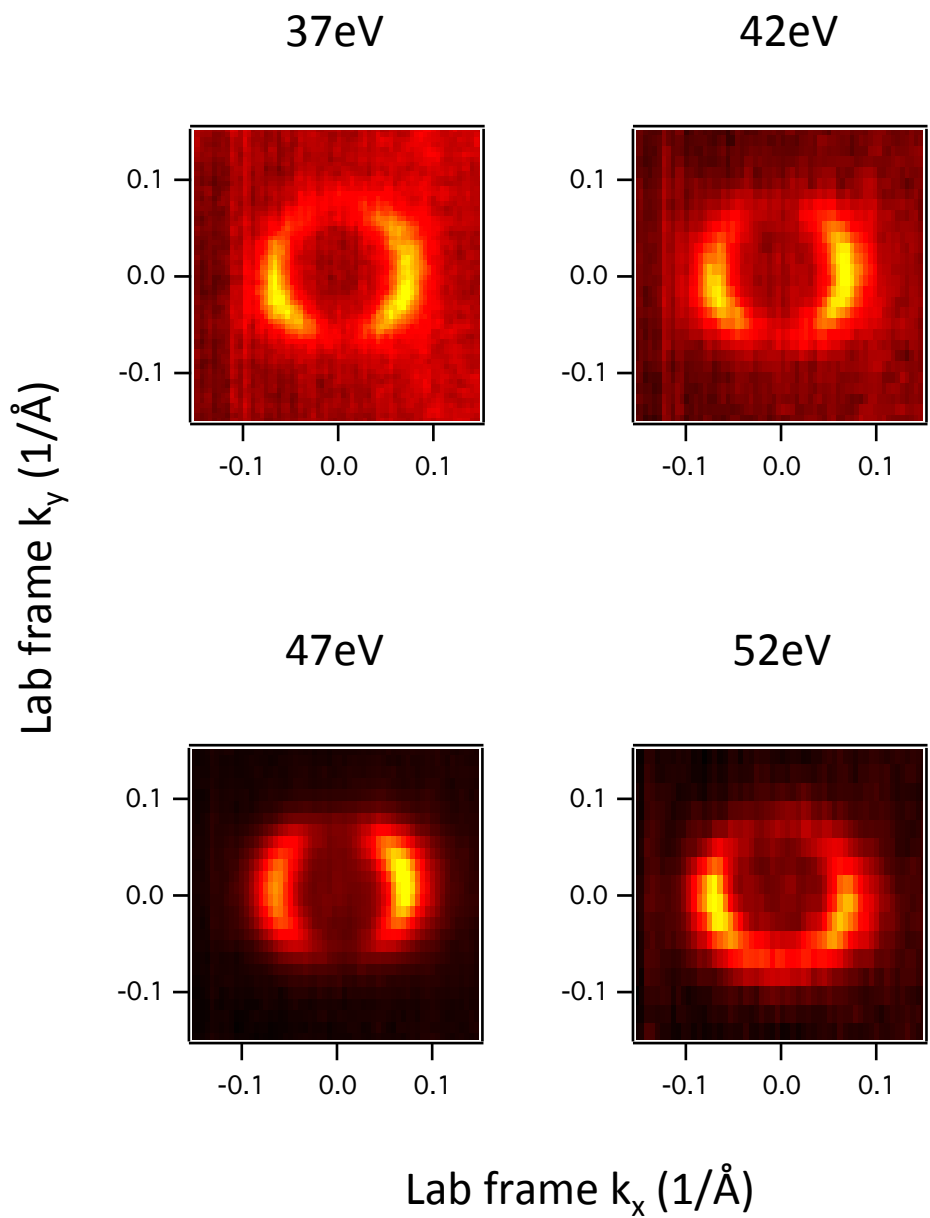

**Figure S3. The constant surface intensity variation of the Dirac state as a function of photon energy.** The plot shows the intensity 200 meV above the Dirac point. The observed intensity variation does not change with photon energy, indicating the final states in the matrix element are free electron states.

**Figure S4. The calculated band structure from a 6 quintuple layer slab of Bi$_2$Se$_3$.** All energies are relative to the Dirac point. The Dirac surface band is represented in the solid green line. The red and blue solid lines are the bulk conduction bands and bulk valence bands, respectively.

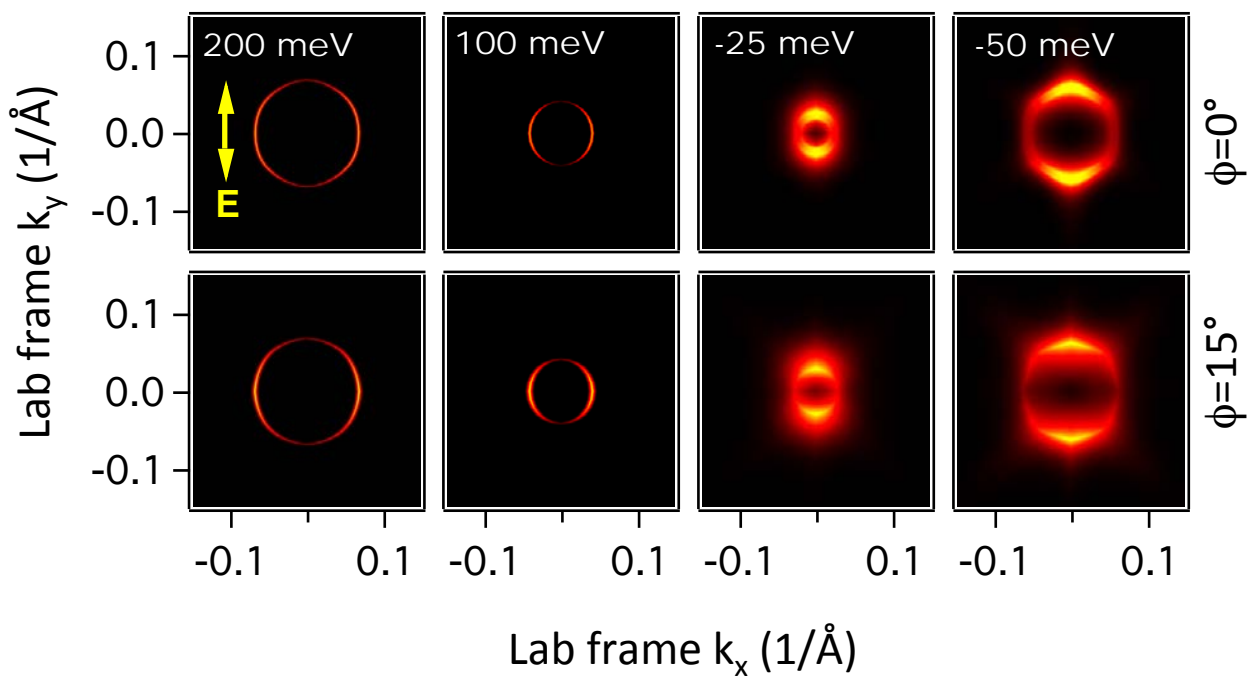

**Figure S5. Calculated constant energy surfaces of a variety of energies for two different sample orientations.** Top panels show original sample orientation with ΓK along $k_x$, and bottom panels have a sample rotation of 15° rotated counterclockwise from the top panels, while leaving the photon polarization fixed to the lab frame. The lack of any change with rotation is consistent with the experimental observations of FIG. 2 b of the main text.

|  | Γ-$k_A$ mirror plane | $\langle\Psi_f|A\cdot p|\Psi_i\rangle$ | Γ-$k_B$ mirror plane |
|---|---|---|---|
| p-polarization | <e\|e\|e> → \|$\Psi_i$> = $p_x$*/$p_z$ | | <e\|e\|e> → \|$\Psi_i$> = $p_y$*/$p_z$ |
| s-polarization | <e\|o\|o> → \|$\Psi_i$> = $p_y$ | | <e\|e\|e> → \|$\Psi_i$> = $p_x$*/$p_z$ |

**Table S1. Initial state orbital wavefunctions with non-vanishing ARPES spectral weight.** The table is shown across two key mirror planes (see Fig 1 of the main text for relevant mirror planes). For * marked states, the spectral weight is much smaller with the momentum / photon polarization combination than the one without the *; but they are not symmetry forbidden.